\documentstyle[aps,epsf,tighten,epsfig,amsmath,multicol]{revtex}
\renewcommand{\narrowtext}{\noindent\begin{multicols}{2}\noindent
\global\columnwidth20.5pc}
\renewcommand{\widetext}{\end{multicols}
\global\columnwidth42.5pc}  
\renewcommand{\top}[1]{%
 \vskip #1%
 \begin{picture}(290,80)(80,500)%
 \thinlines%
 \put(65,500){\line( 1, 0){255}}\put(320,500){\line( 0, 1){5}}%
 \end{picture}%
}
\newcommand{\bottom}[1]{%
 \vskip #1%
 \begin{picture}(290,80)(80,500)%
 \thinlines%
 \put(330,500){\line( 1, 0){255}}\put(330,500){\line( 0, -1){5}}%
 \end{picture}%
}

\def\bi{{\bf i}}
\def\bj{{\bf j}}
\def\bk{{\bf k}}

\def\bq{{\bf q}}

\def\bv{{\bf v}}

\def\b0{{\bf 0}}
\def\cG{{\cal G}}
\def\cO{{\cal O}}
\def\cR{{\cal R}}

\def\cV{{\cal V}}
\def\cZ{{\cal Z}}

\def\bra{\langle}
\def\ket{\rangle}
\def\up{\uparrow}
\def\down{\downarrow}

\def\alf{\alpha}
\def\eps{\epsilon}

\def\Gam{\Gamma}

\def\Lam{\Lambda}

\def\sg{\sigma}

\def\chib{{\bar\chi}}
\def\etab{{\bar\eta}}

\def\psib{{\bar\psi}}

\begin{document}
\title{Renormalization group analysis  of the 2D Hubbard model}
\author{Christoph J. Halboth and Walter Metzner \\
{\em Institut f\"ur Theoretische Physik C, Technische Hochschule Aachen} \\
{\em Templergraben 55, D-52056 Aachen, Germany}}
\date{\small\today}
\maketitle
\begin{abstract}
Salmhofer [Commun.\ Math.\ Phys.\ {\bf 194}, 249 (1998)] has
recently developed a new renormalization group method for
interacting Fermi systems, where the complete flow from the
bare action of a microscopic model to the effective low-energy
action, as a function of a continuously decreasing infrared
cutoff, is given by a differential flow equation which is
local in the flow parameter. We apply this approach to the
repulsive two-dimensional Hubbard model with nearest and
next-nearest neighbor hopping amplitudes.
The flow equation for the effective interaction is evaluated
numerically on 1-loop level. The effective interactions 
diverge at a finite energy scale which is exponentially small
for small bare interactions.
To analyze the nature of the instabilities signalled by the 
diverging interactions we extend Salmhofers renormalization
group for the calculation of susceptibilities. 
We compute the singlet superconducting susceptibilities for
various pairing symmetries and also charge and spin density
susceptibilities.
Depending on the choice of the model parameters (hopping
amplitudes, interaction strength and band-filling) we find
commensurate and incommensurate antiferromagnetic instabilities
or $d$-wave superconductivity as leading instability.
We present the resulting phase diagram in the vicinity of 
half-filling and also results for the density dependence of
the critical energy scale. 

\noindent
\mbox{PACS: 71.10.Fd, 71.10.-w, 74.20.Mn}\\
\end{abstract}

\narrowtext
\section{Introduction}
One of the striking aspects of the high-temperature 
superconducting cuprates is the sensitive dependence of 
their physical properties on the charge carrier concentration 
in the copper-oxide planes, which can be continuously varied 
by doping the inter-plane region. 
In the doping--temperature phase-diagram one generically
finds an antiferromagnetic insulator and a superconducting
phase with $d$-wave symmetry, with strongly doping dependent 
transition temperatures in each case \cite{Gin}. 
\par
The two-dimensional Hubbard model \cite{Mon} is a promising 
prototype model for the electronic degrees of freedom in the
copper-oxide planes. It has an 
antiferromagnetically ordered ground state at half-filling
and is expected to become a $d$-wave superconductor at moderate
doping away from half-filling \cite{Sca}.
\par
Although the Coulomb interaction in the cuprates is certainly
rather strong, there has been considerable recent interest in
the {\em weak}\/ coupling sector of the 2D Hubbard model.
Besides the applicability of (semi-) analytical methods at
weak coupling and the general experience that many strongly
interacting systems are more or less continuously connected
to corresponding weak coupling systems, a major reason for
this interest is that even the weakly interacting 2D Hubbard
model exhibits an extraordinarily rich behavior as a function
of the carrier density and other model parameters.
Conventional perturbation theory breaks down for 
densities close to half-filling, where numerous competing
infrared divergences appear as a consequence of Fermi surface
nesting and van Hove singularities.
These divergencies can in principle be treated by suitable
self-consistent resummations of perturbative contributions to
all orders in the coupling constant. 
Most notably the so-called fluctuation-exchange approximation
\cite{BSW} turned out to be able to describe various expected 
physical properties.
However, a completely unbiased selection of Feynman diagrams
that takes into account all possible particle-particle and
particle-hole channels on equal footing would require at least
the self-consistent summation of all parquet diagrams \cite{DST}, 
which is still beyond present numerical possibilities for 
sufficiently large systems and low temperatures.

Renormalization group (RG) methods are presently
the most promising and best controlled approach to low dimensional
Fermi systems with competing singularities at weak coupling.
Such methods have been developed long ago for one-dimensional
systems where, combined with exact solutions of fixed point
models, they have been a major source of physical insight 
\cite{Sol,Voi}.
Early RG studies of two-dimensional systems started with simple 
but ingenious scaling approaches to the 2D Hubbard model, very 
shortly after the discovery of high-$T_c$ superconductivity 
\cite{Sch1,Dzy1,LMP}. 
These studies focussed on dominant scattering processes between 
van Hove points in k-space, for which a small number of running 
couplings could be defined and computed on 1-loop level. 
Spin-density and superconducting instabilities where identified
from divergencies of the corresponding correlation functions.
Recently, the early scaling approaches have been
revisited by various authors to extract further physical
properties, such as a possible pinning of the Fermi level at the
van Hove singularity \cite{GGV1}, extended saddle points 
\cite{GGV1,Dzy2}, and a possible gap formation on parts of the Fermi 
surface near the van Hove points \cite{FRS}.
Scaling theories with few running couplings have also been used
to analyze instabilities associated with flat Fermi surface 
pieces \cite{HM1,FR} and inflection points on the Fermi surface
\cite{GGV2}.  

A major complication in two-dimensional systems compared to 1D
is that the effective interactions cannot be parameterized
accurately by a 
small number of running couplings, even if irrelevant momentum
and energy dependences are neglected, since the tangential 
momentum dependence of effective interactions along the Fermi 
``surface'' (a line in 2D) is strong and important in the 
low-energy limit. 
This has been demonstrated in particular in a 1-loop RG
study for a model system with two parallel flat Fermi surface 
pieces \cite{ZYD}.
In an impressive series of recent papers Zanchi and Schulz 
\cite{ZS1} have developed a new renormalization group 
approach for interacting Fermi systems, which is
based on a flow equation derived by Polchinski \cite{Pol} in
the context of local quantum field theory.
In this RG version the complete flow from the bare action of
an arbitrary microscopic model to the effective low-energy
action, as a function of a continuously decreasing infrared
energy cutoff, is given by an exact differential flow equation.
Zanchi and Schulz have applied this approach to the 2D Hubbard 
model (with nearest neighbor hopping) in a 1-loop approximation,
with a discretized tangential momentum dependence of the 
effective 2-particle interaction.
The presence of antiferromagnetism and $d$-wave superconductivity 
as major instabilities of the model close to half-filling was 
thereby confirmed.

The development of continuous renormalization group methods
for interacting Fermi systems has made further progress with a
most recent work by Salmhofer \cite{Sal1}.
By expanding the effective action in Wick-ordered monomials
instead of bare monomials he obtained an exact flow equation 
for the effective interactions with a particularly convenient
structure: The $\beta$-function is bilinear in the effective
interactions and {\em local}\/ in the flow parameter, i.e.\ it 
does not depend on effective interactions at higher energy 
scales.

In this work we apply Salmhofer's RG version to the two-dimensional 
Hubbard model
with nearest and also next-nearest neighbor hopping amplitudes,
concentrating on the most interesting electron density regime near
half-filling.
We evaluate the flow of effective (2-particle) interactions on 
1-loop level, neglecting the irrelevant energy dependence and
also the irrelevant normal momentum dependence, but
keeping the important tangential momentum dependence.
As in previous RG calculations, the effective interactions 
diverge at a finite energy scale, which is exponentially small
for a small bare interaction. 
To analyze the physical nature of the instabilities signalled by 
the diverging interactions we extend Salmhofer's RG version for 
the calculation of susceptibilities. 
We compute charge and spin susceptibilities and singlet 
superconducting susceptibilities for various pairing symmetries.
Depending on the choice of the model parameters, hopping 
amplitudes, interaction strength and band-filling, we find
commensurate or incommensurate antiferromagnetic instabilities
or $d$-wave superconductivity as leading instability, in 
qualitative agreement with previous calculations with other
RG versions.
We present the resulting phase diagram of the two-dimensional
Hubbard model near half-filling and results showing how the
critical energy scale decreases away from the half-filled
perfect nesting case. 

We finally note that powerful renormalization group
techniques with a discrete successive reduction of the infrared
cutoff have recently opened the way towards a rigorous 
non-perturbative control of interacting Fermi systems for 
sufficiently small yet finite coupling strength \cite{FMRT}.
Significant rigorous results have already been derived for 2D
systems \cite{FKLT}. 
These mathematical works show in particular that all weak
coupling instabilities in interacting Fermi systems can be
obtained systematically from a renormalization group analysis.

This article is organized as follows. 
In Section II we briefly review Salmhofer's RG for general
interacting Fermi systems, present the explicit flow equations 
for effective 2-particle interactions on 1-loop level, and also 
derive the flow equations for susceptibilities.
In Section III we first discuss how the 1-loop flow equations
are solved, and then present numerical results for the Hubbard
model with a discussion of their physical content.
Section IV closes the presentation with a short summary of
major results and some ideas for further developments.

\section{Renormalization group equations} 
In this section we review Salmhofer's renormalization group
approach for general interacting Fermi systems, present the
explicit flow equations for effective 2-particle interactions
on 1-loop level, and finally derive 1-loop flow equations
for several susceptibilities, which will later be used for our
stability analysis of the 2D Hubbard model. 

\subsection{Functional integral representation} 
We consider a system of interacting spin-$\frac{1}{2}$ fermions
with a single particle basis given by states with a (crystal)
momentum $\bk$, a spin projection $\sg \in \{ \up, \down \}$,
and a kinetic energy $\eps_{\bk}$.
The properties of the system are determined by an action
\begin{equation}
 S[\psi,\psib] = 
 \sum_K (ik_0 - \xi_{\bk}) \, \psib_K \psi_K - V[\psi,\psib]
\end{equation}
where $K = (k_0,\bk,\sg)$ is a multi-index containing the Matsubara
frequency $k_0$ in addition to the single particle quantum numbers;
$\psib_K$ and $\psi_K$ are Grassmann variables associated with 
creation and annihilation operators,
$\xi_{\bk} = \eps_{\bk} - \mu$ is the single-particle energy 
relative to the chemical potential, 
and $V[\psi,\psib]$ is an arbitrary many-body interaction.
The non-interacting single-particle propagator of the system 
is given by
\begin{equation}
 C(K) = \frac{1}{ik_0 - \xi_{\bk}}
\end{equation}
All connected Green functions can be obtained from the generating
functional \cite{NO}
\begin{equation}
 \cG[\eta,\etab] = 
 \log \left\{ \int\! d\mu_C[\psi,\psib] \, 
 e^{-V[\psi,\psib]} \, e^{(\psib,\eta) + (\etab,\psi)}
 \right\}
\end{equation}
with the normalized Gaussian measure
\begin{equation}\begin{split}
 d\mu_C[\psi,\psib] &= \\
    \prod_K& d\psi_K d\psib_K \, e^{(\psib,C^{-1}\psi)} \Big/
 \int\prod_K d\psi_K d\psib_K \, e^{(\psib,C^{-1}\psi)}
\end{split}
\end{equation}
Here we have introduced the short-hand notation
$(\chi,\psi) = \sum_K \chi_K \psi_K$ and
$(C^{-1}\psi)_K = C^{-1}(K) \, \psi_K$ 
for arbitrary Grassmann variables $\chi_K$ and $\psi_K$.
Note the identity
\begin{equation}
 \int\! d\mu_C[\psi,\psib] \, e^{(\psib,\eta) + (\etab,\psi)} =
 e^{-(\etab,C\eta)}
\end{equation}
which implies that $\cG[\eta,\etab] = - (\etab,C \eta)$ in the
non-interacting case $V[\psi,\psib] = 0$. 
The connected $m$-particle Green functions are given by
\begin{align}
 G_m(K'_1,\dots,&K'_m;K_1,\dots,K_m) = \nonumber\\ 
 &
 (-1)^m \bra \psi_{K'_1} \dots \psi_{K'_m} 
             \psib_{K_m} \dots \psib_{K_1} \ket_c 
 \nonumber \\
 =\,& \left.
 \frac{\partial^m}{\partial\etab_{K'_1} \dots \partial\etab_{K'_m}}
 \frac{\partial^m}{\partial\eta_{K_m} \dots \partial\eta_{K_1}}
 \cG[\eta,\etab] \right|_{\eta = \etab = 0}
\end{align}
where $\bra \dots \ket_c$ is the connected average of the 
product of Grassmann variables between the brackets.
\par
The renormalization group equations are most conveniently
formulated for another generating functional, the 
{\em effective interaction} \cite{Sal2}
\begin{equation}
 \cV[\chi,\chib] = 
 - \log \left\{ \int\! d\mu_C[\psi,\psib] \, 
 e^{-V[\psi+\chi,\psib+\chib]} \right\}
\end{equation}
A simple substitution of variables yields the relation
\begin{equation}
 - \cV[\chi,\chib] = (\etab,C \eta) + \cG[\eta,\etab] \quad
 \mbox{where} \quad \chi = C \eta, \, \chib = C \etab
\end{equation}
Hence, functional derivatives of $\cV[\chi,\chib]$ generate 
connected Green functions divided by 
$C(K_1) \dots C(K_m) \, C(K'_1) \dots C(K'_m)$,
i.e.\ \mbox{(non-interacting)} propagators are amputated from external 
legs in the corresponding Feynman diagrams. 
The term $(\etab,C \eta)$ cancels the non-interacting part of
$\cG[\etab,\eta]$ such that $\cV[\chi,\chib] = 0$ for 
$V[\psi,\psib] = 0$.
Hence, the non-interacting propagator is subtracted from the
one-particle Green function generated by $\cV$.

\subsection{Wick-ordered continuous RG}
We now briefly review the derivation of a continuous 
renormalization group equation for Wick-ordered (amputated) 
Green functions, as first derived in the context of interacting
Fermi systems by Salmhofer \cite{Sal1}. 

The original system is endowed with an infrared cutoff at an
energy scale $\Lam > 0$ by replacing the bare propagator $C(K)$ 
with
\begin{equation}
 C^{\Lam}(K) = \frac{\Theta^{\Lam}_{>}(K)}{ik_0 - \xi_{\bk}}
\end{equation}
Here $\Theta^{\Lam}_{>}(K)$ is a function that vanishes for
$\sqrt{k_0^2 + \xi_{\bk}^2} \ll \Lam$ and tends to one for
$\sqrt{k_0^2 + \xi_{\bk}^2} \gg \Lam$. 
In this way the infrared singularity of the propagator at 
$k_0 = 0$ and $\xi_{\bk} = 0$ (corresponding to the 
non-interacting Fermi surface in $\bk$-space) is cut off at
scale $\Lam$.
A simple choice for $\Theta^{\Lam}_{>}(K)$, which we will later
adopt in our numerical calculations, is
\begin{equation}
 \Theta^{\Lam}_{>}(K) = \Theta(|\xi_{\bk}| - \Lam)
\end{equation}
where $\Theta$ is the usual step function. With this choice
single-particle states close to the Fermi surface are strictly
excluded from the theory.
Alternatively, one may also use a smooth cutoff function.

The generating functional for connected Green functions and
the effective interaction constructed with $C^{\Lam}$ (instead
of $C$) will be denoted by $\cG^{\Lam}[\eta,\etab]$ and 
$\cV^{\Lam}[\chi,\chib]$, respectively. 
The original functionals $\cG$ and $\cV$ are recovered in the 
limit $\Lam \to 0$.
It is not hard to show that the effective interaction satisfies
the following exact {\em renormalization group equation} 
\cite{Sal1,BW}
\widetext
\top{-2.8cm}
\begin{equation}\label{rge}
 \frac{\partial}{\partial\Lam} \cV^{\Lam}[\chi,\chib] =
   \sum_K \dot{C}^{\Lam}(K) \,
        \frac{\partial^2 \cV^{\Lam}[\chi,\chib]}
        {\partial\chi_K \, \partial\chib_K}
 - \sum_K \dot{C}^{\Lam}(K) \,
        \frac{\partial\cV^{\Lam}[\chi,\chib]}{\partial\chi_K} 
     \, \frac{\partial\cV^{\Lam}[\chi,\chib]}{\partial\chib_K}
\end{equation}
\bottom{-2.7cm}
\narrowtext
where $\dot{C}^{\Lam}(K) = \partial C^{\Lam}(K)/\partial\Lam$.
With the initial condition 
\begin{equation}
 \cV^{\Lam_0}[\chi,\chib] = V[\chi,\chib]
\end{equation}
this equation determines the flow of $\cV^{\Lam}$ uniquely for 
all $\Lam < \Lam_0$.
The initial value $\Lam_0$ must be chosen such that 
$\Theta^{\Lam}_{>}(K) = 0$ for all $K$ and $\Lam > \Lam_0$.
For the step function cutoff introduced above, $\Lam_0$ is the 
maximal value of $|\xi_{\bk}|$.

An expansion of the functional $\cV^{\Lam}[\chi,\chib]$ in the 
renormalization group equation (\ref{rge}) in powers of $\chi_K$
and $\chib_K$ would lead to Polchinski's \cite{Pol} flow 
equations for amputated connected Green functions, which have
been applied to the 2D Hubbard model by Zanchi and Schulz~\cite{ZS1}.
Alternatively, one can also expand with respect to 
{\em Wick-ordered}\/ monomials 
%
\begin{equation}\label{wickexp}\begin{split}
 \cV^{\Lam}[&\chi,\chib] = \sum_{m=0}^{\infty}
 \frac{1}{(m!)^2} \sum_{K_1,\dots,K_m} \sum_{K'_1,\dots,K'_m} \times
 \\ 
& \times V^{\Lam}_{m}(K'_1,\dots,K'_m;K_1,\dots,K_m) 
\; e^{-\Delta_{D^{\Lam}}} 
 \prod_{j=1}^m \chib_{K'_j} \chi_{K_j}
\end{split}
\end{equation}
%
The exponent in the Wick-ordering operator is the differential 
operator
\begin{equation}
 \Delta_{D^{\Lam}} = \sum_K D^{\Lam}(K) \,
 \frac{\partial^2}{\partial\chi_K \, \partial\chib_K}
\end{equation}
with the propagator 
\begin{equation}
 D^{\Lam}(K) = C(K) - C^{\Lam}(K) = 
 \frac{\Theta^{\Lam}_{<}(K)}{ik_0 - \xi_{\bk}}
\end{equation} 
where $\Theta^{\Lam}_{<}(K) = 1 - \Theta^{\Lam}_{>}(K)$.
Note that $D^{\Lam}$ contributes in the infrared regime 
excluded from $C^{\Lam}$.
The Wick-ordered monomials reduce to pure monomials in the
limit $\Lam \to 0$, since $D^{\Lam}(K) \to 0$ in that limit.
Hence, the functions $V^{\Lam}_m$ tend to the usual amputated
connected Green functions for $\Lam \to 0$.

Inserting the expansion (\ref{wickexp}) for 
$\cV^{\Lam}[\chi,\chib]$ on the left hand side
of the RG equation (\ref{rge}) yields two terms, with the 
$\Lam$-derivation acting on the coefficients $V^{\Lam}_m$ or
on the Wick-ordered monomials, respectively. Since
\begin{equation}\begin{split}
 \frac{\partial}{\partial\Lam} \, 
 e^{-\Delta_{D^{\Lam}}} & \prod_{j=1}^m \chib_{K'_j} \chi_{K_j} = \\
 & - \sum_K \dot{D}^{\Lam}(K) \,
 \frac{\partial^2}{\partial\chi_K \, \partial\chib_K} \, 
 e^{-\Delta_{D^{\Lam}}} \prod_{j=1}^m \chib_{K'_j} \chi_{K_j}
\end{split}
\end{equation}
and $\dot{D}^{\Lam} = - \dot{C}^{\Lam}$,
the second term on the left hand side cancels the first term 
on the right hand side of (\ref{rge}).
Only the second term, quadratic in $\cV^{\Lam}$, remains.
Expanding this term with respect to Wick monomials and 
comparing coefficients, one can express the $\Lam$-derivative
of $V^{\Lam}_m$ as a {\em bilinear}\/ form of all the other 
functions $V^{\Lam}_{n}$ \cite{Sal1}. 
A graphical representation of these flow equations is shown
in Fig.\ 1. 

The precise general equation for $V^{\Lam}_m$ has rather 
complicated combinatorial factors and will not be written here, 
since we will compute only $V^{\Lam}_2$ in a 1-loop 
approximation.
Note that one internal line in Fig.\ 1 corresponds to 
$\dot{D}^{\Lam}$ and the others to the {\em low}\/ energy 
propagator $D^{\Lam}$. 
Note also that the flow equations are {\em local}\/ in $\Lam$.

\subsection{1-loop equations}
To detect dominant instabilities of the system in the weak
coupling limit, it is sufficient to truncate the infinite 
\begin{figure}[t]
\epsfxsize8.5cm
\centering\leavevmode\epsfbox{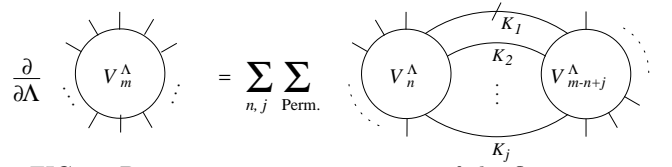}
\caption{Diagrammatic representation of the flow equation
 for the Wick-ordered amputated Green-functions $V^{\Lam}_m$. 
 The internal line with a slash corresponds to $\dot{D}^{\Lam}$,
 the others to $D^{\Lam}$; all possible pairings leaving $m$ 
 ingoing and $m$ outgoing external legs have to be summed.}
\label{fig1}
\end{figure}
\begin{figure}[t]
\epsfxsize8.5cm
\centering\leavevmode\epsfbox{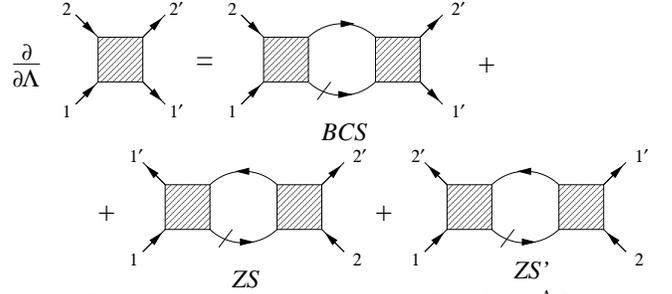}
\caption{Flow equation for the vertex function 
 $\Gam^{\Lam}$ in 1-loop approximation with the particle-particle
 channel (BCS) and the two particle-hole channels (ZS and ZS').}
\label{fig2}
\end{figure}
\noindent
hierarchy
of flow equations described by Fig.\ 1 at {\em one-loop}\/ 
level, 
and neglect all components of the effective interaction except the
two-particle interaction $V^{\Lam}_2$. 
Since self-energy corrections are also neglected, $V^{\Lam}_2$
reduces to the one-particle irreducible two-particle {\em vertex
function}\/, and will therefore be denoted by $\Gam^{\Lam}$ from
now on.
Summing all possible pairings of two vertices in a one-loop
diagram as in Fig.\ 2, 
one obtains the {\em flow equation}
\begin{equation}\label{1loop}
 \renewcommand{\arraystretch}{1.3}
 \begin{split}
 \frac{\partial}{\partial\Lam} 
 \Gam^{\Lam}(K'_1&,K'_2; K_1,K_2) = \\
 \, & \frac{1}{\beta V} \sum_{K,K'}
 \frac{\partial}{\partial\Lam} \left[
  D^{\Lam}(K) \, D^{\Lam}(K') \right] \\
 & \times  \Big[ 
  \frac{1}{2} 
    \Gam^{\Lam}(K'_1,K'_2;K,K') \, \Gam^{\Lam}(K,K';K_1,K_2)
 \\ & \quad
  - \Gam^{\Lam}(K'_1,K;K_1,K') \, \Gam^{\Lam}(K',K'_2;K,K_2)
 \\ & \quad
  + \Gam^{\Lam}(K'_2,K;K_1,K') \, \Gam^{\Lam}(K',K'_1;K,K_2)
   \Big]
 \end{split}
\end{equation}
where $\beta$ is the inverse temperature and $V$ is the volume
of the system.
The three terms on the right hand side are the contributions
from the Cooper (or BCS) channel and the two zero-sound channels
(ZS and ZS').
Note that for translation invariant systems momentum 
conservation implies that $\Gam^{\Lam}(K'_1,K'_2;K_1,K_2) \neq 0$
only if $k_1 + k_2 = k'_1 + k'_2$, so that the sum over $k$ and 
$k'$ in (\ref{1loop}) is reduced to a single energy-momentum sum.

For a spin-rotation invariant system the {\em spin structure}\/ of 
the vertex function can be written as
\begin{equation}\begin{split}
 \Gam^{\Lam}(K'_1,K'_2;K_1,K_2) = \,&
 \Gam^{\Lam}_s(k'_1,k'_2;k_1,k_2) \, 
 S_{\sg'_1,\sg'_2;\sg_1,\sg_2} +\\
&+ \Gam^{\Lam}_t(k'_1,k'_2;k_1,k_2) \, 
 T_{\sg'_1,\sg'_2;\sg_1,\sg_2} 
\end{split}
\end{equation}
where 
\begin{equation}
\begin{split}  
 S_{\sg'_1,\sg'_2;\sg_1,\sg_2} = 
 \textstyle \frac{1}{2} \, \left(
 \delta_{\sg_1\sg'_1}\delta_{\sg_2\sg'_2} -
 \delta_{\sg_1\sg'_2}\delta_{\sg_2\sg'_1} \right) 
 \\
 T_{\sg'_1,\sg'_2;\sg_1,\sg_2} = 
 \textstyle \frac{1}{2} \, \left(
 \delta_{\sg_1\sg'_1}\delta_{\sg_2\sg'_2} +
 \delta_{\sg_1\sg'_2}\delta_{\sg_2\sg'_1} \right)
\end{split}
\end{equation}
are the projection operators on singlet and triplet states
in a two-particle spin space, respectively.
The antisymmetry of $\Gam^{\Lam}$ with respect to 
$K_1 \leftrightarrow K_2$ or $K'_1 \leftrightarrow K'_2$
implies that $\Gam^{\Lam}_s$ is symmetric and $\Gam^{\Lam}_t$
antisymmetric under exchange of the variables $k_1$ and $k_2$
or $k'_1$ and $k'_2$.
Carrying out the spin sum in the flow equation one obtains
%
\begin{equation}
\begin{split}
 \partial_{\Lam}& \Gam^{\Lam}_{\alf}(k'_1,k'_2;k_1,k_2) =\\
 \,& \sum_{i=s,t} \sum_{j=s,t} \big[
 C^{\rm BCS}_{\alf ij} \beta^{\rm BCS}_{ij}(k'_1,k'_2;k_1,k_2) +\\
&\quad + C^{\rm ZS}_{\alf ij} \beta^{\rm ZS}_{ij}(k'_1,k'_2;k_1,k_2) +
 C^{\rm ZS'}_{\alf ij} \beta^{\rm ZS'}_{ij}(k'_1,k'_2;k_1,k_2)
 \big]
\end{split}
\end{equation}
%
for $\alf = s,t$, where the coefficients $C_{\alf ij}$ can be
grouped in matrices
\begin{equation}
 \renewcommand{\arraystretch}{1.5}
 \begin{array}{lcl}
 C^{\rm BCS}_s = 
 \left( {1 \atop 0} \, {0 \atop 0} \right) & \quad &
 C^{\rm ZS}_s = - C^{\rm ZS'}_s = \frac{1}{4}
 \left( {-1 \atop 3} \, {3 \atop 3} \right) \\
 C^{\rm BCS}_t = 
 \left( {0 \atop 0} \, {0 \atop 1} \right) & \quad &
 C^{\rm ZS}_t = C^{\rm ZS'}_t = \frac{1}{4}
 \left( {1 \atop 1} \, {1 \atop 5} \right) \end{array}
\end{equation}
and the ``$\beta$-functions'' are given by
\begin{equation}
 \renewcommand{\arraystretch}{1.5}
 \begin{array}{rl}
 \beta^{\rm BCS}_{ij}(k'_1,k'_2;k_1,k_2) =&
  \displaystyle \frac{1}{2\beta V} \sum_{k,k'} \partial_{\Lam} 
  \left[ D^{\Lam}(k) \, D^{\Lam}(k') \right] \times \\
  &  \times
  \Gam^{\Lam}_i(k'_1,k'_2;k,k') \, \Gam^{\Lam}_j(k,k';k_1,k_2) \\
 \beta^{\rm ZS}_{ij}(k'_1,k'_2;k_1,k_2) =&
  - \displaystyle \frac{1}{\beta V} \sum_{k,k'} \partial_{\Lam} 
  \left[ D^{\Lam}(k) \, D^{\Lam}(k') \right] \times \\
  &  \times
  \Gam^{\Lam}_i(k'_1,k;k_1,k') \, \Gam^{\Lam}_j(k',k'_2;k,k_2) \\
 \beta^{\rm ZS'}_{ij}(k'_1,k'_2;k_1,k_2) =&
 - \beta^{\rm ZS}_{ij}(k'_2,k'_1;k_1,k_2)
 \end{array}
\end{equation}

We finally list some useful relations for the vertex function
following from general {\em symmetries}.
Time-reversal invariance implies
\begin{equation}\begin{split}
 \Gam^{\Lam}(K'_1,K'_2;K_1,K_2) \,=\, &
 \sg'_1 \sg'_2 \sg_1 \sg_2 \, \times \\
& \times \Gam^{\Lam}(\cR K_1,\cR K_2;\cR K'_1,\cR K'_2)
\end{split}
\end{equation}
where $\cR K = (k_0,-\bk,-\sg)$ for $K = (k_0,\bk,\sg)$, and 
the number $1$ $(-1)$ is assigned to $\sg = \,\up$ $(\down)$
in the prefactor.
Assuming in addition spatial reflection invariance and spin
rotation invariance, one obtains
\begin{equation}
 \Gam^{\Lam}(K'_1,K'_2;K_1,K_2) = \Gam^{\Lam}(K_1,K_2;K'_1,K'_2)
\end{equation}
From the behavior under complex conjugation
\begin{equation}
 {\bar\Gam}^{\Lam}(K'_1,K'_2;K_1,K_2) =
 \Gam^{\Lam}({\bar K}_1,{\bar K}_2;{\bar K}'_1,{\bar K}'_2)
\end{equation}
with ${\bar K} = (-k_0,\bk,\sg)$ one can then deduce that the
vertex function $\Gam^{\Lam}(K'_1,K'_2;K_1,K_2)$ is real, if 
all the energy variables vanish.

\subsection{Susceptibilities}
To identify possible instabilities of the system we compute
various susceptibilities, i.e.\ the linear response of the
system to small external fields.

Application of an external field $h$ leads to an additional
contribution to the action    
\begin{equation}
 V'[h;\psi,\psib] = - \sum_q \, 
 \left[ {\bar h}(q) O(q) + h(q) {\bar O}(q) \right]
\end{equation}
where $O(q)$ is a bilinear form in the Grassmann variables,
${\bar O}(q)$ its hermitian conjugate, and ${\bar h}(q)$ is 
the complex conjugate of $h(q)$.
We will compute the response to fields coupling to the 
{\em charge density}
\begin{equation}
 \rho(q) = \sum_{k,\sg} \psib_{k-q,\sg} \psi_{k,\sg}
\end{equation}
and the {\em spin density}\/ in $z$-direction
\begin{equation}
 s^z(q) = \sum_k \left[ \psib_{k-q,\up} \psi_{k,\up} - 
   \psib_{k-q,\down} \psi_{k,\down} \right]
\end{equation}
and also the response to pairing fields coupling to the 
{\em singlet pair}\/ operator
\begin{equation}
 \Delta(q) = \sum_{k,\sg} d(\bk+\bq/2) \,
 \psi_{k+q,\up} \psi_{-k,\down}
\end{equation}
where $d(\bk)$ is a function with even parity specifying the
orbital symmetry of the pairing operator ($s$-wave, $d$-wave etc.).
The linear response of the expectation value
\begin{equation}
 m(q) = V^{-1} \bra O(q) \ket = 
 \frac{1}{V} \, \frac{\partial \log \cZ[h]}
                     {\partial {\bar h}(q)}
\end{equation}
with the partition function
\begin{equation}
 \cZ[h] = 
 \int \! d\mu_C[\psi,\psib] \, 
 e^{-V[\psi,\psib] - V'[h;\psi,\psib]}
\end{equation}
is given by the susceptibility
\begin{equation}
 \chi(q) = 
 \left. \frac{\partial m(q)}{\partial h(q)} \right|_{h=0} =
 V^{-1} \! \left. \bra O(q) {\bar O}(q) \ket \right|_{h=0}
\end{equation}
We consider only systems which are translation invariant,
spin-rotation invariant, and charge conserving in the absence
of the external field $h$.
In the normal (not symmetry-broken) phase the expectation
values $\bra O(q) \ket$ then vanish for $h(q) \to 0$, except 
the expectation value $\bra \rho(0) \ket$, which yields the 
average particle number of the system.

The renormalization group equation (\ref{rge}) is not affected 
by the presence of the additional term $V'$ in the bare action, 
since an arbitrary many-body interaction was allowed anyway.
Only the initial condition of the flow is modified to
\begin{equation}
 \cV^{\Lam_0}[h;\chi,\chib] = 
 V[\chi,\chib] + V'[h;\chi,\chib]
\end{equation}
which leads to an $h$-dependent effective interaction
$\cV^{\Lam}[h;\chi,\chib]$.
When a pairing field is coupled to the system, the expansion
of $\cV^{\Lam}[h;\chi,\chib]$ with respect to Wick-ordered
monomials will also contain monomials where the number of 
creation and annihilation variables is not equal, and
Eq.\ (\ref{wickexp}) must be generalized to
\widetext
\top{-2.8cm}
\begin{equation}
 \cV^{\Lam}[h;\chi,\chib] =
 \sum_{m,n=0}^{\infty} 
 \frac{1}{m!n!} \sum_{K'_1,\dots,K'_m} \sum_{K_1,\dots,K_n}
 V^{\Lam}_{mn}([h];K'_1,\dots,K'_m;K_1,\dots,K_n) \;
 e^{-\Delta_{D^{\Lam}}} \,
 \chib_{K'_1} \dots \chib_{K'_m} \, 
 \chi_{K_n} \dots \chi_{K_1}
\end{equation}
\bottom{-2.7cm}
\narrowtext

To obtain the {\em linear}\/ response of the system, we expand the 
effective $m$-body interactions $V^{\Lam}_m$ (or its generalization
$V^{\Lam}_{mn}$) in powers of $h$.
The effective 0-body interaction can be expanded as
\begin{equation}
 V^{\Lam}_0[h] =  
 V^{\Lam}_0 - V \sum_q \chi^{\Lam}(q) \, {\bar h}(q) \, h(q) 
 + \cO(h^3)
\end{equation}
Note that $V^{\Lam}_0[h]$ converges to the grand-canonical
potential $\Omega[h] = - \log\cZ[h]$ for $\Lam \to 0$,
and $\chi^{\Lam}(q)$ converges to the susceptibility $\chi(q)$.
If $h$ couples to the charge density, there is also a linear
contribution 
$- N^{\Lam} [ h(0) + {\bar h}(0) ]$ 
in the expansion of $V^{\Lam}_0[h]$, where $N^{\Lam}$ converges 
to the average particle number for $\Lam \to 0$.
For a field coupling to charge or spin density, the effective 
1-body interaction becomes
\begin{equation}\begin{split}
 V^{\Lam}_1([h];K';&K) = 
 V^{\Lam}_1(K';K) - \\
&- \sum_q \left[ R^{\Lam}(q;K';K) \, {\bar h}(q) 
  \; + \; {\rm h.c.} \right]  + \cO(h^2)
\end{split}
\end{equation}
The spin structure of the renormalized vertex $R^{\Lam}$ is
\begin{equation}
 R^{\Lam}(q;K';K) = \delta_{\sg'\sg} \, R^{\Lam}_C(q;k';k)
\end{equation}
for the charge vertex and
\begin{equation}
 R^{\Lam}(q;K';K) = 
 \sg \, \delta_{\sg'\sg} \, R^{\Lam}_S(q;k';k)
\end{equation}
for the spin vertex.
For a pairing field, $V^{\Lam}_1$ has only quadratic terms in
$h$, but the off-diagonal effective interactions
\begin{equation}
 \renewcommand{\arraystretch}{1.3}
 \begin{array}{rl}
 V^{\Lam}_{0,2}([h];K_1,K_2) =& \displaystyle
 - 2 \sum_q R^{\Lam}(q;K_1,K_2) \, {\bar h}(q) + \cO(h^2) \\
 V^{\Lam}_{2,0}([h];K'_1,K'_2) =& \displaystyle
 - 2 \sum_q {\bar R}^{\Lam}(q;K'_1,K'_2) \, h(q) + \cO(h^2) \\
 \end{array}
\end{equation}
have linear (and higher order) contributions.
The spin structure of the renormalized singlet pairing vertex 
is
\begin{equation}
 R^{\Lam}(q;K_1,K_2) = 
 \sg_1 \, \delta_{\sg_1,-\sg_2} \, R^{\Lam}_s(q;k_1,k_2)
\end{equation}
Effective two- and many-body interactions can be expanded
similarly. 
Inserting the expansions of the effective interactions 
$V^{\Lam}_{mn}$ into 
the flow equations and comparing the coefficients of 
contributions with equal powers in the external field, one
obtains flow equations for these coefficients, especially
for $\chi^{\Lam}$ and $R^{\Lam}$.

We will again truncate the infinite hierarchy of flow
equations at 1-loop level, and neglect (zero-field) self-energy
terms by setting $V^{\Lam}_1(K';K) = 0$.
To obtain the linear response susceptibility it is then
sufficient to solve the two flow equations for the susceptibility
$\chi^{\Lam}$ and the renormalized charge, spin, or pairing vertex 
$R^{\Lam}$ represented diagrammatically in Fig.\ 3.
If $h$ couples to the charge or spin density, these equations
read
%
\begin{equation}\begin{split}
 \frac{\partial}{\partial\Lam} \chi^{\Lam}(q) \,=\, &
 \frac{1}{\beta V} \sum_{K,K'} 
 \partial_{\Lam} \left[ D^{\Lam}(K) D^{\Lam}(K') \right] \, \times\\
& \times  R^{\Lam}(q;K';K) \, R^{\Lam}(-q;K;K')
\end{split}
\end{equation}
and
\begin{equation}\begin{split}
 \frac{\partial}{\partial\Lam} R^{\Lam}(q;K'_1;K_1) \,=\, &
 - \frac{1}{\beta V} \sum_{K,K'} 
 \partial_{\Lam} \left[ D^{\Lam}(K) D^{\Lam}(K') \right] \, \times \\
&\times\,  R^{\Lam}(q;K';K) \, \Gam^{\Lam}(K'_1,K;K_1,K')
\end{split}
\end{equation}
with the initial condition 
$R^{\Lam_0}(q;K';K) = \delta_{\sg'\sg} \delta_{k-k',q}$
for the charge and
$R^{\Lam_0}(q;K';K) = \sg\delta_{\sg'\sg} \delta_{k-k',q}$
for the spin vertex part.
Note that $R^{\Lam}(q;K';K) \neq 0$ only if $k'=k-q$.
If $h$ couples to pair creation and annihilation operators,
the flow equations become
\begin{equation}\begin{split}
 \frac{\partial}{\partial\Lam} \chi^{\Lam}(q) =&
 - \frac{2}{\beta V} \sum_{K_1,K_2} 
 \partial_{\Lam} \left[ D^{\Lam}(K_1) D^{\Lam}(K_2) \right] \,\times \\
&\times\, R^{\Lam}(q;K_1,K_2) \, {\bar R}^{\Lam}(q;K_1,K_2)
\end{split}
\end{equation}
\begin{figure}
\begin{minipage}{\columnwidth}
\epsfxsize7cm
\centering\leavevmode\epsfbox{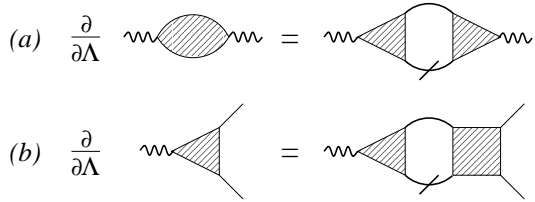}
\caption{Flow equations for (a) the susceptibilities
 $\chi^{\Lam}$ 
\mbox{and (b)} the response vertices $R^{\Lam}$ in 1-loop 
 approximation.}
\label{fig3}
\end{minipage}
\end{figure}
\noindent and
\begin{equation}\begin{split}
 \frac{\partial}{\partial\Lam} R^{\Lam}(q;K_1,K_2) \,=\,&
 \frac{1}{2\beta V} \sum_{K'_1,K'_2} 
 \partial_{\Lam} \left[ D^{\Lam}(K'_1) D^{\Lam}(K'_2) \right] \,\times\\
&\times\, R^{\Lam}(q;K'_1,K'_2) \, \Gam^{\Lam}(K'_1,K'_2;K_1,K_2)
\end{split}
\end{equation}
%
with the initial condition
$R^{\Lam_0}(q;K_1,K_2) = 
\sg_1 \delta_{\sg_1,-\sg_2} \delta_{k_1+k_2,q} \, 
d \left( \frac{\bk_1 - \bk_2}{2} \right)$.
Note that \\ $R^{\Lam}(q;K_1,K_2) \neq 0$ only if $k_1+k_2=q$.
The initial condition for the susceptibility is 
$\chi^{\Lam_0}(q) = 0$ in all cases.

The vertex function $\Gam^{\Lam}$ is obtained by solving the
flow equations (\ref{1loop}) in the absence of a perturbing
field. One can then solve the linear differential equation for
$R^{\Lam}$ and finally integrate the flow equation for 
$\chi^{\Lam}$.
Note that for the special case of a flat fermi surface these 1-loop
equations have the same structure as the  
so-called fast parquet equations \cite{ZYD}, but here the flow 
variable is the cutoff $\Lam$ instead of an external energy or 
momentum variable. 

\section{Application to the 2D Hubbard model}
We will now apply the general renormalization group method 
derived in the previous section to the two-dimensional Hubbard
model, the main aim being an analysis of the leading instabilities 
of the system at weak coupling.

\subsection{Hubbard model and Fermi surface}
The Hubbard model \cite{Mon} is a lattice electron model
with the simple Hamiltonian
\begin{equation}\label{HM}
 H = \sum_{\bi,\bj} \sum_{\sg} t_{\bi\bj} \,
 c^{\dag}_{\bi\sg} c_{\bj\sg} +
 U \sum_{\bj} n_{\bj\up} n_{\bj\down} \; ,
\end{equation}
where $c^{\dag}_{\bi\sg}$ and $c_{\bi\sg}$ are the usual
creation and annihilation operators for fermions with spin
projection $\sg \in \{ \up,\down \}$ on a lattice site 
${\bi}$.
We consider the Hubbard model with a repulsive interaction
$U>0$ on a (two-dimensional) square lattice with a 
hopping amplitude 
\begin{equation}
 t_{\bi\bj} = \left\{ \begin{array}{rl}
 -t  & \mbox{if $\bi$ and $\bj$ are nearest neighbors} \\
 -t' & \mbox{if $\bi$ and $\bj$ are next-nearest neighbors} \\
  0  & \mbox{else} \end{array} \right.
\end{equation}
\begin{figure}
\centering\leavevmode\epsfbox{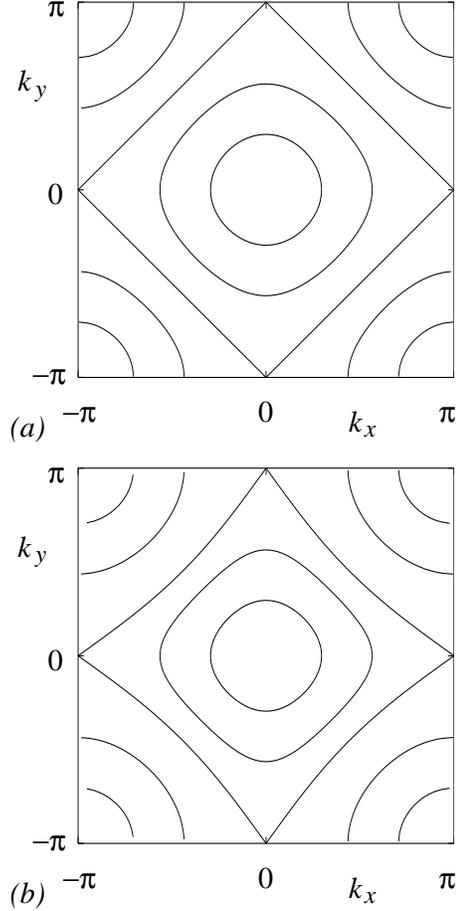}\vspace{5mm}
\caption{The Fermi surfaces of the non-interacting 2D 
 Hubbard model with $t' = 0$ (a) and $t' = - 0.16t$ (b) for 
 various choices of the chemical potential $\mu$.}
\label{fig4}
\end{figure}
\noindent 
The corresponding dispersion relation of non-interacting 
particles is
\begin{equation}
 \eps_{\bk} = -2t(\cos k_x + \cos k_y)
              -4t'(\cos k_x \, \cos k_y)
\end{equation}
This dispersion relation has saddle points at $\bk = (0,\pi)$
and $(\pi,0)$, which lead to logarithmic van Hove 
singularities in the non-interacting density of states at the 
energy $\eps = \eps_{vH} = 4t'$.
In Fig.\ 4 we show the Fermi surfaces of the non-interacting
system for various electron densities $n$ for the case without 
next-nearest neighbor hopping ($t'=0$) and a case with $t'<0$.

For a chemical potential $\mu = \eps_{\rm vH}$ the Fermi 
surface contains the van Hove points.
In this case a perturbative calculation of the two-particle 
vertex function leads to several infrared divergencies already 
at second order in $U$, i.e.\ at 1-loop level (see the Feynman
diagrams in Fig.\ 2) \cite{Sch1,Dzy1,LMP}.
In particular,
the particle-particle channel diverges as $\log^2$ for vanishing 
total momentum $\bk_1+\bk_2$, and logarithmically for 
$\bk_1+\bk_2 = (\pi,\pi)$.
The particle-hole channel diverges logarithmically for vanishing
momentum transfer; for momentum transfer $(\pi,\pi)$ it diverges
logarithmically if $t' \neq 0$ and as $\log^2$ in the special 
case $t'=0$.
Note that $\mu=0$ for $t'=0$ corresponds to half-filling ($n=1$).
For $t'=0$ there are also logarithmic divergences for all
momentum transfers parallel to $(\pi,\pi)$ or $(\pi,-\pi)$
due to the strong nesting of the square shaped Fermi surface.
For $\mu \neq \eps_{\rm vH}$ only the usual logarithmic Cooper 
singularity at zero total momentum in the particle-particle 
channel remains.
However, the additional singularities at $\mu = \eps_{\rm vH}$ 
lead clearly to largely enhanced contributions for small 
$\mu - \eps_{\rm vH}$, especially if $t'$ is also 
small.

Hence, for small $\mu - \eps_{\rm vH}$ one has to deal with 
competing divergencies in different channels.
This problem can be treated systematically by the RG method
described in Sec.\ II.

\subsection{Parameterization of the vertex function}
We now prepare for a computation of the 1-loop flow of the
vertex function and susceptibilities.
The flow equations derived in Sec.\ II cannot be solved 
analytically.
Even a numerical solution is not possible with the full
energy and momentum dependence of the vertex function,
which must therefore be suitably simplified.

The dependence of the vertex function on energy variables can be 
neglected completely without much damage, because it is absent in
the bare interaction, and irrelevant (in the sense of power
counting) in the low-energy limit (see, for example, Ref.\ 
\cite{Sha}).
We therefore approximate
\begin{equation}
 \Gam^{\Lam}_{\alf}(k'_1,k'_2;k_1,k_2) \approx
 \Gam^{\Lam}_{\alf}(\bk'_1,\bk'_2;\bk_1,\bk_2)
\end{equation}
Choosing an energy independent cutoff function 
$\Theta^{\Lam}_{<}(\bk)$, the Matsubara sums on the right hand 
side of the flow equations can then be performed analytically.
One thus obtains
\widetext
\top{-2.8cm}
\begin{equation}
\begin{split}
 \partial_{\Lam} 
 \Gam^{\Lam}_{\alf}(\bk'_1,\bk'_2;& \bk_1,\bk_2) =   \\ 
 & \sum_{i=s,t} \sum_{j=s,t} \left[
 C^{\rm BCS}_{\alf ij} 
 \beta^{\rm BCS}_{ij}(\bk'_1,\bk'_2;\bk_1,\bk_2) +
 C^{\rm ZS}_{\alf ij} 
 \beta^{\rm ZS}_{ij}(\bk'_1,\bk'_2;\bk_1,\bk_2) +
 C^{\rm ZS'}_{\alf ij} 
 \beta^{\rm ZS'}_{ij}(\bk'_1,\bk'_2;\bk_1,\bk_2)
 \right]
\end{split}
\end{equation}
for $\alf = s,t$, where the $\beta$-functions are now frequency
independent and read
\begin{equation}
\begin{split}
 \beta^{\rm BCS}_{ij}(\bk'_1,\bk'_2;\bk_1,\bk_2) =&
  \displaystyle \frac{1}{2V} \sum_{\bk,\bk'} 
  \partial_{\Lam} 
  \left[ \Theta^{\Lam}_{<}(\bk) \, \Theta^{\Lam}_{<}(\bk') \right]
  \frac{f(-\xi_{\bk}) - f(\xi_{\bk'})}{\xi_{\bk} + \xi_{\bk'}} \,
  \Gam^{\Lam}_i(\bk'_1,\bk'_2;\bk,\bk') \, 
  \Gam^{\Lam}_j(\bk,\bk';\bk_1,\bk_2) \\
 \beta^{\rm ZS}_{ij}(\bk'_1,\bk'_2;\bk_1,\bk_2) =&
  - \displaystyle \frac{1}{V} \sum_{\bk,\bk'} 
  \partial_{\Lam} 
  \left[ \Theta^{\Lam}_{<}(\bk) \, \Theta^{\Lam}_{<}(\bk') \right]
  \frac{f(\xi_{\bk}) - f(\xi_{\bk'})}{\xi_{\bk} - \xi_{\bk'}} \,
  \Gam^{\Lam}_i(\bk'_1,\bk;\bk_1,\bk') \, 
  \Gam^{\Lam}_j(\bk',\bk'_2;\bk,\bk_2) \\
 \beta^{\rm ZS'}_{ij}(\bk'_1,\bk'_2;\bk_1,\bk_2) =&
 - \beta^{\rm ZS}_{ij}(\bk'_2,\bk'_1;\bk_1,\bk_2)
\end{split}
\end{equation}
\bottom{-2.7cm}
\narrowtext
with the Fermi function 
$f(\xi) = \left[ e^{\beta\xi} + 1 \right]^{-1}$.
Note that momentum conservation implies that $\bk$ and $\bk'$ 
are related by $\bk + \bk' = \bk_1 + \bk_2$ in the
Cooper channel and by $\bk + \bk'_1 = \bk' + \bk_1$ in the
zero sound channel, such that only one independent momentum 
variable has to be summed.

For a step cutoff function 
$\Theta^{\Lam}_{<}(\bk) = \Theta(\Lam - |\xi_{\bk}|)$
one has
\begin{equation}\label{deltatheta}
 \partial_{\Lam} 
 \left[ \Theta^{\Lam}_{<}(\bk) \, \Theta^{\Lam}_{<}(\bk') \right]
 = \delta(\Lam - |\xi_{\bk}|) \, \Theta(\Lam - |\xi_{\bk'}|) +
 \bk \leftrightarrow \bk'
\end{equation}
such that the $\bk$-sum can be reduced to a one-dimensional 
integral in the thermodynamic limit.
To this end we substitute $k_x$ and $k_y$ by the new variables
$\xi = \xi_{\bk}$ and the angle $\phi$ between $\bk$ and the
$x$-axis in $\bk$-space, and use
\begin{equation}\begin{split}
 \frac{1}{V} \sum_{\bk}& \dots \to \\
& \int \frac{dk_x}{2\pi} \, \frac{dk_y}{2\pi} \dots =
 \frac{1}{(2\pi)^2} \int d\xi \int_0^{2\pi} d\phi \, 
  J(\xi,\phi) \dots
\end{split}
\end{equation}
where $J(\xi,\phi)$ is the Jacobian associated with the
transformation of variables.
Since the integrand contains a factor $\delta(\Lam - |\xi|)$,
the $\xi$-integration can be performed by hand, leaving only
the angular integral over $\phi$ to be done numerically.

The flow equation can be solved only if also the momentum
dependence of the vertex function is simplified.
At least for weak coupling (in practice also for moderate
ones), the vertex function acquires strong momentum dependences 
only for momenta close to the Fermi surface. 
Note that for the Hubbard model the bare vertex function 
$\Gam^{\Lam_0}$ does not depend on momentum at all.
Instabilities are signalled by divergencies of the vertex
function $\Gam^{\Lam}$ for momenta close to the Fermi surface
and small $\Lam$.
Hence we will focus on the flow of the vertex function with 
momenta close to the Fermi surface.

The intermediate momenta $\bk$ and $\bk'$ are constrained to a 
momentum shell around the Fermi surface, since 
$|\xi_{\bk}|,|\xi_{\bk'}| \leq \Lam$ (see Eq.\ (\ref{deltatheta})).
Hence, the values of the vertex function at momenta within that 
$\Lam$-shell govern their own flow.

For fixed finite $\Lam$, the dependence of 
$\Gam^{\Lam}_{\alf}(\bk'_1,\bk'_2;\bk_1,\bk_2)$ on $\xi_{\bk_1}$
etc.\ is regularized by the cutoff for 
$|\xi_{\bk_1}|,\dots < \Lam$. 
For momenta within the $\Lam$-shell one may therefore approximate
the vertex function by
\begin{equation}\label{projection}
 \Gam^{\Lam}_{\alf}(\bk'_1,\bk'_2;\bk_1,\bk_2) \approx
 \Gam^{\Lam}_{\alf}
  (\bk'_{F1},\bk_{F1} + \bk_{F2} - \bk'_{F1};\bk_{F1},\bk_{F2})
\end{equation}
where $\bk_{F1}$ etc.\ are projections of $\bk_1$ etc.\ on
the Fermi surface (see Fig.\ 5).
Note that strong momentum dependences of the effective vertex are 
built up only by contributions with intermediate momenta $\bk$ and 
$\bk'$ (on the right hand side of the flow equations) which are close 
to the Fermi surface, because for such momenta the fractions 
$\frac{f(\mp\xi_{\bk}) - f(\xi_{\bk'})}
 {\xi_{\bk} \pm \xi_{\bk'}}$
can be big. Hence, for the most important momenta, the error made 
by the projection is relatively small (even if $\Lam$ is not small), 
because these momenta are close to their projected counterparts. 
The projected vertex function can be parameterized by three angles
$\phi_1,\phi_2,\phi_3$ associated with $\bk_{F1}$, $\bk_{F2}$ and 
$\bk'_{F1}$, respectively, i.e.
\begin{equation}
 \Gam^{\Lam}_{\alf}
 (\bk'_{F1},\bk_{F1} + \bk_{F2} - \bk'_{F1};\bk_{F1},\bk_{F2}) =
 \Gam^{\Lam}_{\alf}(\phi_1,\phi_2,\phi_3)
\end{equation}
The angular dependence turns out to be strong for small $\Lam$ 
and cannot be neglected. 

With the above projection procedure only 
functional dependences which are irrelevant in the low-energy
limit have been neglected (see, for example, Ref.\ \cite{Sha}).
The approximation (\ref{projection}) is asymptotically exact for
$\Lam \to 0$ (for momenta within the $\Lam$-shell)
and, for the Hubbard model, also for $\Lam = \Lam_0$.
At intermediate stages of the flow there are of course corrections.
To assess their importance, we have refined the parameterization of 
the vertex function in some test runs by using a second surface of 
constant energy in momentum space as a target for momentum 
projections.
In addition to the Fermi surface a canonical choice is the 
{\em van Hove surface}\/ defined by $\eps_{\bk} = \eps_{\rm vH}$,
where $\eps_{\rm vH}$ is the energy at which the density of states
has a van Hove singularity due to saddle points of $\eps_{\bk}$.
Momenta are then projected on that surface (Fermi or van Hove)
which is energetically closer.
In this way scattering processes between van Hove points, which
are particularly important at scales 
$\Lam = |\eps_{\rm vH} - \mu|$, are treated more accurately.
In addition to the three angles $\phi_1,\phi_2,\phi_3$ one needs
three binary variables $\nu_1,\nu_2,\nu_3$ to label the
closest projection surface for $\bk_1$, $\bk_2$ and $\bk'_1$.

The parameterization of the vertex parts $R^{\Lam}$ required for 
the computation of susceptibilities is done in a similar fashion.
We concentrate on {\em static}\/ susceptibilities ($q_0 = 0$)
and neglect the (irrelevant) energy dependence 
of $R^{\Lam}$.
For the charge and spin density vertex we approximate
\begin{equation}
 R^{\Lam}_{C,S}(\bq;\bk';\bk) \approx
 R^{\Lam}_{C,S}(\bq;\bk_F \!-\! \bq;\bk_F)
\end{equation}
where $\bk_F$ is the projection of $\bk$ on the Fermi surface.
The pairing susceptibility is computed only for the most 
\begin{figure}
\epsfxsize6cm
\centering\leavevmode\epsfbox{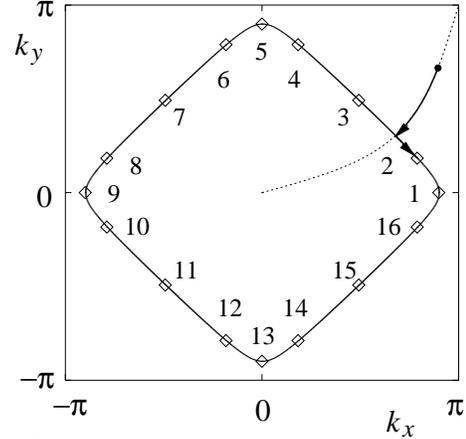}
\caption{Projection of momenta on the Fermi surface;
 discretization and labelling of angle variables.}
\label{fig5}
\end{figure}
\noindent
interesting case of vanishing total momentum $\bq = 0$.
The corresponding vertex part is approximated by
\begin{equation}
 R^{\Lam}_s(\b0;\bk,-\bk) \approx 
 R^{\Lam}_s(\b0;\bk_F,-\bk_F)
\end{equation}
The projection approximation for the vertex parts is again exact
for $\Lam = \Lam_0$ and asymptotically exact for $\Lam \to 0$.
It can also be improved for intermediate $\Lam$ by projecting 
momenta either on the Fermi surface or on the van Hove surface, 
whichever is closer.

The behavior of the {\em compressibility}\/ 
$\chi_C = \partial n/\partial\mu = \chi_C(\b0)$ and the 
{\em homogeneous spin susceptibility}\/ $\chi_S = \chi_S(\b0)$
cannot be obtained directly from the flow equations for
$\chi^{\Lam}_{C,S}(\bq)$.
The problem is that $\chi^{\Lam}_{C,S}(\b0)$ vanishes for all 
$\Lam > 0$ (at zero temperature), since the infrared cutoff 
blocks particle-hole excitations with an infinitesimal momentum 
transfer.
We will therefore compute $\chi_C$ and $\chi_S$ from the
effective quasi-particle interaction 
\begin{equation}
 f^{\sg\sg'\Lam}_{\bk_F\bk'_F} = 
 Z^{\Lam}_{\bk_F} Z^{\Lam}_{\bk'_F} \,
 \Gam^{\Lam}(\bk_F\sg,\bk'_F\sg';\bk_F\sg,\bk'_F\sg')
\end{equation}
Note that the forward scattering limit (zero momentum
and energy transfer) of the two-particle vertex is unique for 
$\Lam > 0$ and converges to the quasi-particle interaction
for $\Lam \to 0$ and $\bk_F \neq \bk'_F$ \cite{MCD}.
The wave function renormalization factor $Z^{\Lam}_{\bk_F}$
is one in our calculation because we have neglected 
self-energy contributions.
Following the usual Fermi liquid arguments \cite{PN}
one obtains the compressibility and the homogeneous spin 
susceptibility as
\begin{equation}
 \chi_{C,S} = 2 \int \frac{d^2k}{(2\pi)^2} \,
 X^{C,S}_{\bk_F} \, \delta(\eps_{\bk} - \mu)
\end{equation}
where $X^{C,S}_{\bk_F}$ is the solution of the inhomogeneous 
linear integral equation
\begin{equation}
 X^{C,S}_{\bk_F} + 
 2 \int \frac{d^2k'}{(2\pi)^2} \, f^{C,S}_{\bk_F\bk'_F} \,
 X^{C,S}_{\bk'_F} \, \delta(\eps_{\bk'} - \mu) = 1
\end{equation}
with $f^{C,S}_{\bk_F\bk'_F} = 
\frac{1}{2} \left( f^{\sg\sg}_{\bk_F\bk'_F} \pm 
f^{\sg,-\sg}_{\bk_F\bk'_F} \right)$.
In a non-interacting system one would obtain 
$X^{C,S}_{\bk_F} = 1$.
The quantity
\begin{equation}
 \frac{\partial s_{\bk_F}}{\partial\mu} 
 = \frac{1}{|\bv_{\bk_F}|} \, X^C_{\bk_F}
\end{equation}
with the velocity $\bv_{\bk} = \nabla_{\bk} \eps_{\bk}$
describes the linear response of the Fermi surface in $\bk_F$
(a shift along the normal vector) for a small shift of the 
chemical potential.

For a concrete numerical solution of the flow equations the
angular dependence of the vertex function is discretized, with 
a finer mesh in the vicinity of the saddle points of $\eps_{\bk}$ 
at $(\pi,0)$ and $(0,\pi)$, as shown in Fig.\ 5. 
One is thus left with a finite (though very large) number of
flowing coupling constants. 
If not stated otherwise, we have used 16 points on the Fermi
surface to discretize the vertex function, corresponding to
$672$ independent (i.e.\ not symmetry-related) coupling constants, if
the fermi level lies at the van~Hove energy $\epsilon_{vHS}$, and 
$846$ couplings otherwise.
With the addition of 16 points on the van Hove surface one has
to deal with $4728$ flowing couplings.

\subsection{Results}
We have computed the flow of the vertex function and the
susceptibilities for several choices of the bare interaction 
$U>0$, the next-nearest neighbor hopping amplitude $t' \leq 0$ 
and the chemical potential $\mu$, where $t'$ and $\mu$ have
been chosen such that the Fermi surface is on or close to the
van Hove points of $\eps_{\bk}$, and the particle density is
close to half-filling.

In all cases the vertex function develops a strong momentum
dependence for small $\Lam$ with divergencies for several
momenta at some critical scale $\Lam_c > 0$, which vanishes
exponentially $U \to 0$.
To see which physical instability is associated with the
diverging vertex function we have computed the following
susceptibilities: 

\begin{itemize}
\item[i)] 
commensurate antiferromagnetic spin susceptibility 
$\chi_S(\pi,\pi)$,
\item[ii)] 
incommensurate spin susceptibility $\chi_S(\bq)$ with
$\bq = (\pi-\delta,\pi)$ and $\bq = (1-\delta) (\pi,\pi)$
\cite{Sch90},
\item[iii)] 
commensurate charge susceptibility $\chi_C(\pi,\pi)$ ,
\item[iv)] 
singlet pair susceptibilities with form factors \cite{Sca}
\begin{equation}
 d(\bk) =  \left\{ \begin{array}{ll}
 1 & 
 \mbox{($s$-wave)} \\     
 \frac{1}{\sqrt{2}} (\cos k_x + \cos k_y) &
 \mbox{(extended $s$-wave)} \\
 \frac{1}{\sqrt{2}} (\cos k_x - \cos k_y) &
 \mbox{($d$-wave $d_{x^2-y^2}$)} \\
 \sin k_x \sin k_y &
 \mbox{($d$-wave $d_{xy}$)}. \end{array} \right.
\end{equation}
\end{itemize}
Some of these susceptibilities diverge together with the vertex
function at the scale $\Lam_c$.
Depending on the choice of $U$, $t'$ and $\mu$ the strongest
divergence is found for the commensurate or incommensurate
spin susceptibility or for the pair susceptibility with 
$d_{x^2-y^2}$ symmetry. 

We will now present explicit results for the flow of the
two-particle vertex and susceptibilities for a coupling
strength $U = t$, which is much smaller than the bandwidth
$W = 8t$ and therefore safely in the weak 
coupling regime.
All energy scales will be plotted in units of $t$.
To exhibit the interaction-induced renormalizations of the
susceptibilities, we plot the flow of the ratio 
$\chi^{\Lam}/\chi_0^{\Lam}$, where $\chi_0^{\Lam}$ is the  
susceptibility of the non-interacting system at scale $\Lam$, 
as obtained from the flow equations for $U=0$. 
We show examples for the flow of $\chi_0^{\Lam}$ in the Appendix. 
Note that the non-interacting susceptibilities $\chi_0^{\Lam}$ are 
all finite for $\Lam > 0$, such that a divergence of $\chi^{\Lam}$
at a finite scale $\Lam_c$ implies a diverging ratio
$\chi^{\Lam}/\chi_0^{\Lam}$ and vice versa.

In Fig.\ 6 we show the flow for $t'=0$ and $\mu = -0.005$,
corresponding to a density $n = 0.995$, i.e.\ almost at 
half-filling. 
Here and in the following we plot the singlet part of the
vertex function for a selected choice of momenta on the
Fermi surface, including those momenta for which the vertex
function renormalizes most strongly.
The singlet vertex function has its largest values for 
um\-klapp scattering along the diagonal of the Brillouin zone, 
but also forward and Cooper scattering of particles on
opposite sides of the almost square Fermi surface are 
strongly enhanced. Scattering amplitudes for momenta near
the van Hove points diverge a bit more slowly.
The triplet part of the vertex function is renormalized 
mostly for forward and Cooper scattering, but generally
more weakly than the singlet part.  
The spin susceptibility with an antiferromagnetic wave vector
clearly dominates over pairing susceptibilities in this case.
The incommensurate spin susceptibilities are indistinguishable
from the commensurate one in Fig.\ 6 because the incommensurability
parameter $\delta$ is almost zero so close to half-filling
(see Ref.\ \cite{Sch90}).
Note also that the susceptibility ratios for isotropic and
extended $s$-wave pairing are equal here, and almost coincide 
with the charge density susceptibility ratio.
The non-interacting susceptibility for extended $s$-wave pairing
(and thus $\chi$)
is however much smaller than the other two (see Appendix).

Decreasing the density (away from half-filling) one enters
a regime where pairing correlations with $d_{x^2-y^2}$ 
symmetry dominate at sufficiently low energy scales. 
This is seen in Fig.\ 7, where we show the flow for $t'=0$
and $\mu = -0.02$, corresponding to $n = 0.984$.
Note that for small $U$ the transition from antiferromagnetism
to superconductivity occurs already at a critical density $n_c$ 
quite close to half-filling. For increasing $U$ the deviation of
$n_c$ from half-filling increases.
The flow in Fig.\ 7 exhibits a threshold at $\Lam=2|\mu|$ below
which the amplitudes for various scattering processes,
especially umklapp scattering, renormalize only very slowly. 
The flow of the antiferromagnetic spin 
susceptibility is cut off at the same scale.
The infinite slope singularity in some of the flow curves at
scale $\Lam=|\mu|$ is due to the van Hove singularity  being
\begin{figure}
\centering\leavevmode\epsfbox{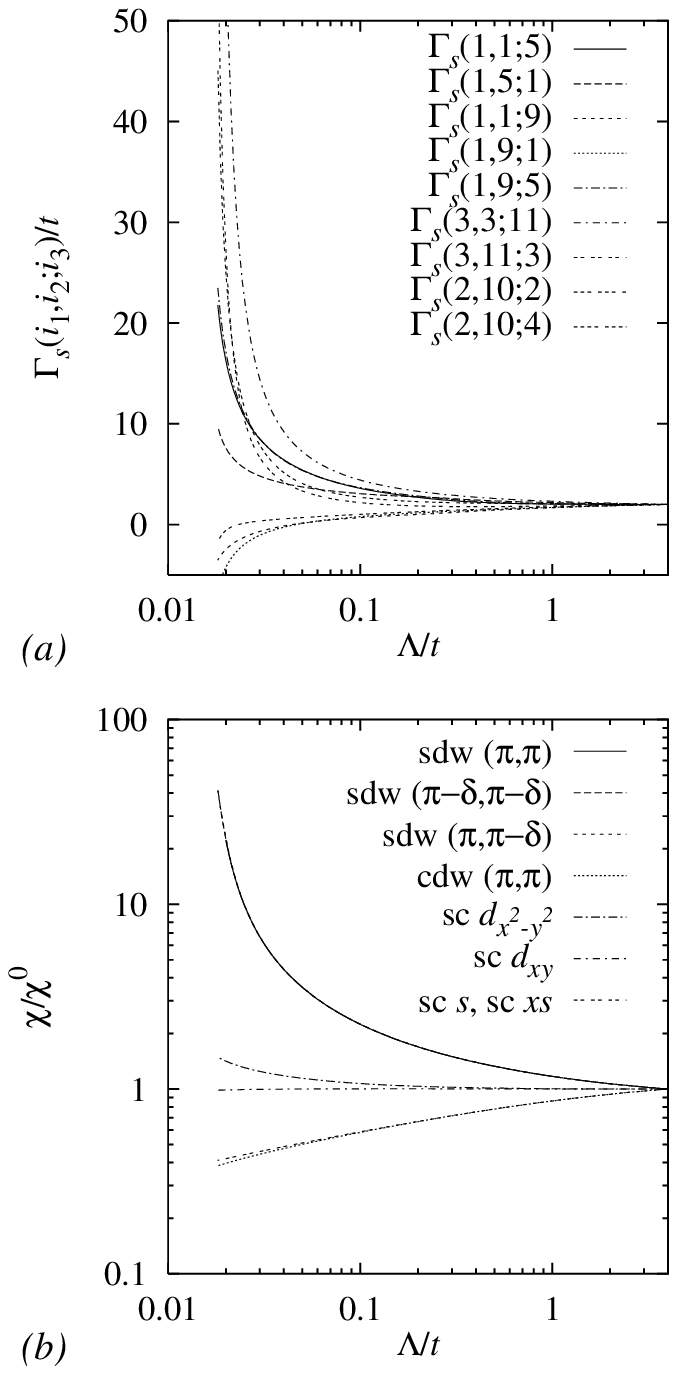}
\caption{(a) The flow of the singlet vertex function 
 $\Gam^{\Lam}_s$ as a function of $\Lam$ for several choices of 
 the momenta $\bk_{F1}$, $\bk_{F2}$ and $\bk'_{F1}$, which are
 labelled according to the numbers in Fig.\ 5. 
 The model parameters are $U = t$ and $t'=0$, the chemical 
 potential $\mu = -0.005$;
 (b) the flow of the ratio of interacting and non-interacting 
 susceptibilities, $\chi^{\Lam}/\chi_0^{\Lam}$ for the same 
 system.}
\label{fig6}
\end{figure}
\noindent
crossed at that scale. 
The pairing susceptibility with $d_{x^2-y^2}$-symmetry is
obviously dominant here (note the logarithmic scale).
Following the flow of the vertex function and susceptibilities
one can see that the $d_{x^2-y^2}$-pairing correlations
develop in the presence of pronounced but 
{\em short-range antiferromagnetic spin-correlations}, 
in agreement with earlier ideas on $d$-wave superconductivity 
\cite{Sca}.

In Fig.\ 8 we show the $(\mu,U)$ phase diagram for $t'=0$ 
obtained by identifying the dominant instability from the flow 
for many different values of $\mu$ and $U$.
Note that $\mu = 0$ corresponds to half-filling.
The regime with a leading commensurate antiferromagnetic
spin density instability is separated from the $d$-wave pairing
regime by 
\begin{figure}
\centering\leavevmode\epsfbox{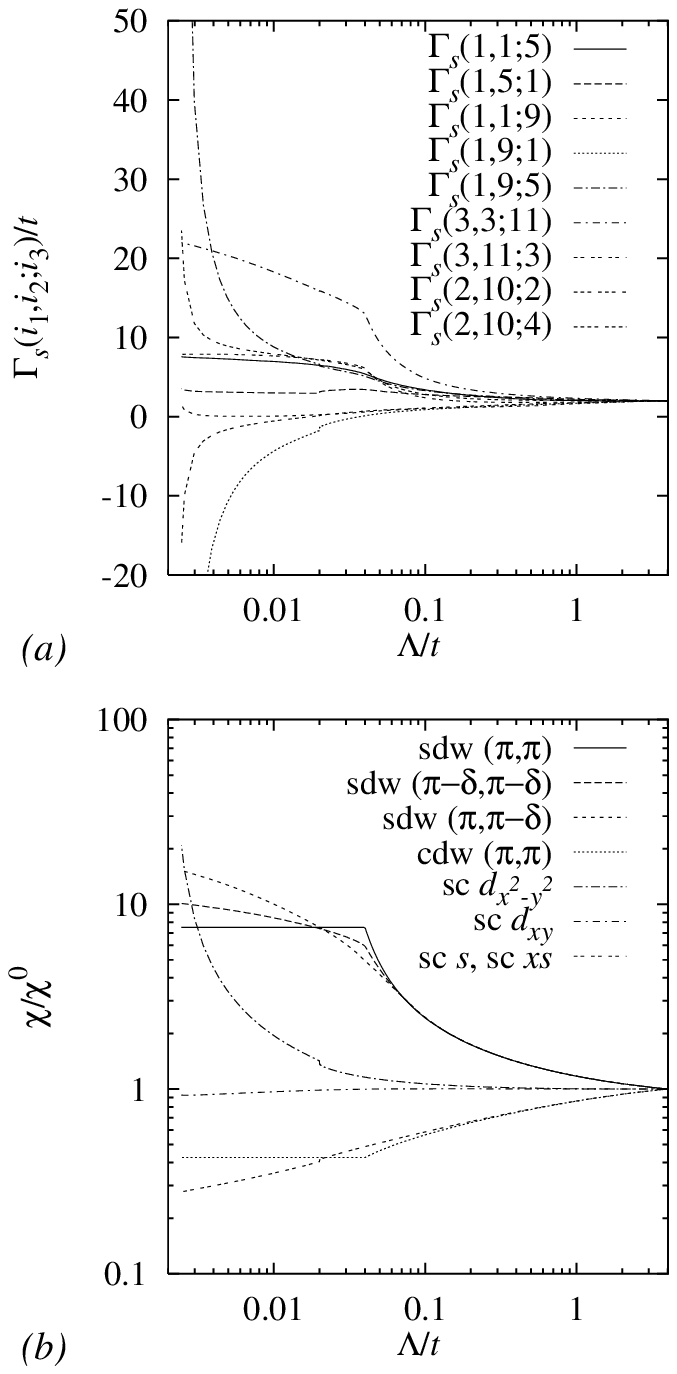}
\caption{Same as Fig.\ 6 for $U = t$, $t' = 0$ and 
 $\mu = -0.02$.}
\label{fig7}
\end{figure}
\noindent
a thin region where incommensurate spin density
fluctuations with $\bq=(\pi,\pi-\delta)$ dominate.
Other incommensurate structures may be more favorable than the ones
considered here.
For $U \to 0$ at fixed density $n < 1$ the
superconducting instability always dominates, because the bare
particle-hole bubbles are finite away from half-filling, while
the Cooper channel always diverges at least logarithmically.
The way the critical energy scale $\Lam_c$ varies as the system is
doped away from half-filling can be seen in Fig.\ 9 for an
interaction strength $U = t$.

The different symbols show which instability is leading at 
$\Lam_c$. The two straight lines represent the linear functions
$\Lam_c = |\mu|$ and $\Lam_c = 2|\mu|$, respectively.
As already observed by Zanchi and Schulz \cite{ZS1}, the 
superconducting instability is leading if $\Lam_c < |\mu|$.
This may be related to the fact that only pair fluctuations receive a
singular enhancement at $\Lam = |\mu|$ (see Fig.\ 7), while 
spin fluctuations don't. 
A commensurate spin density wave instability cannot be favorable
for $\Lam_c < 2|\mu|$, since their 
\begin{figure}
\centering\leavevmode\epsfbox{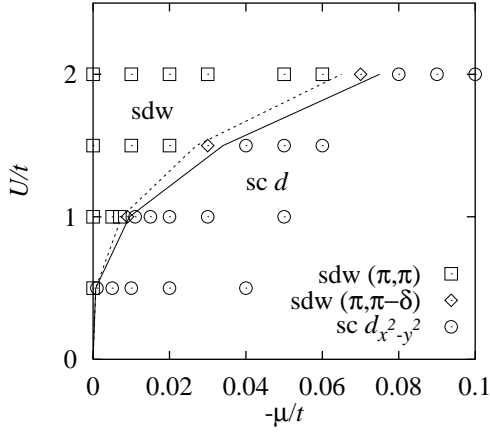}
\caption{The $(\mu,U)$ phase diagram for $t'=0$ near half-filling;
the symbols represent the parameters for which the flow has been 
computed; the solid line separates the spin-density wave regime
from the superconducting regime, the dotted line separates the 
commensurate and incommensurate spin-density regions.}
\label{fig8}
\end{figure}
\begin{figure}
\centering\leavevmode\epsfbox{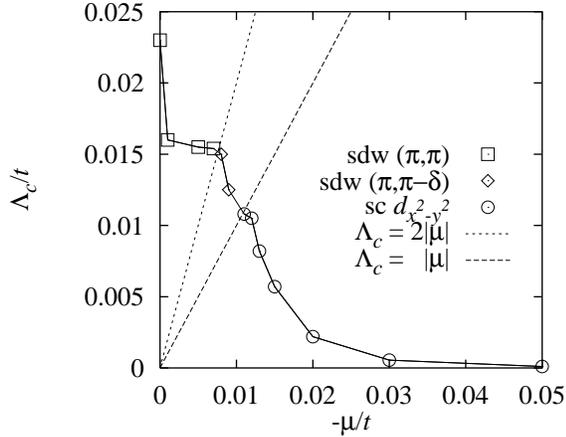}
\caption{The critical energy scale $\Lam_c$ as a function of
 the chemical potential $\mu$ for $U = t$ and $t'=0$. 
The different symbols indicate, whether the leading instability is a
commensurate or incommensurate spin-density wave or $d$-wave pairing
instability; 
 the straight lines represent the functions $\Lam_c = |\mu|$
 and $\Lam_c = 2|\mu|$, respectively.}
\label{fig9}
\end{figure}
\noindent
 flow  is cut off at $\Lam = 2|\mu|$
(see Fig.\ 7 once again). Hence, the in\-com\-men\-surate
 spin density wave is the leading for
$|\mu| < \Lam_c < 2|\mu|$.
The sharp peak in $\Lam_c$ at $\mu=0$ (half-filling) is due to
the van Hove singularity.
For larger deviations from half-filling, the critical energy scale
$\Lam_c$ vanishes rapidly. 
Note, however, that for larger values of $U$ the regime with a 
sizable scale $\Lam_c$ extends to larger values of $\mu$, i.e.\ 
to larger doping.

In Fig.\ 10 (a) and (b) we show results for the compressibility and 
the homogeneous spin susceptibility, respectively, for two  
choices of the chemical potential, at $U = t$ and $t'=0$.
We recall that these quantities have~been~ob-
\begin{figure}
\centering\leavevmode\epsfbox{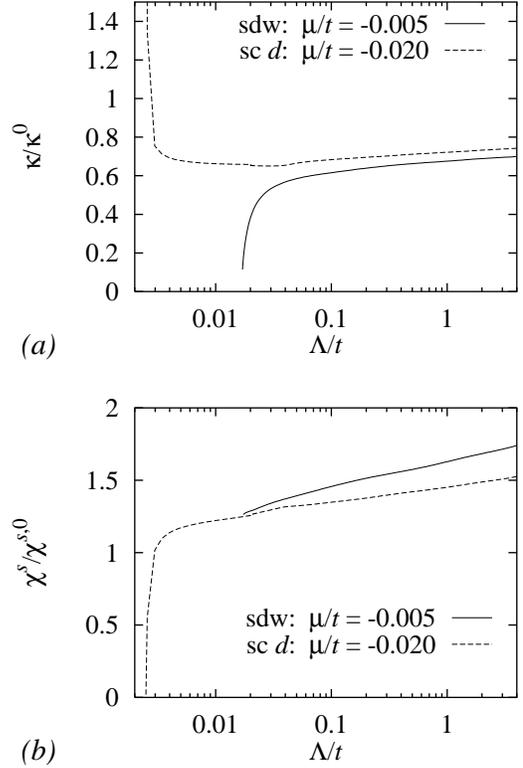}
\caption{The flow of (a) the compressibility and (b) the homogeneous 
 spin susceptibility as a 
 function of $\Lam$ for various choices of $\mu$ at $U = 1$ and
 $t'=0$; $\kappa^0$ and $\chi^{s,0}$ are the corresponding 
 non-interacting quantities.}
\label{fig10}
\end{figure}
\noindent
tained from the
forward scattering vertex by using Fermi liquid relations as
discussed above.
The non-interacting compressibility $\kappa^0$ and spin
susceptibility $\chi^{s,0}$ in the plotted ratios are defined
without infrared cutoff. 
Hence the flow in Fig.\ 10 is entirely due to the flow of the Landau
function, starting at the simple RPA result for the Hubbard model 
at $\Lam=\Lam_0$.
Close to half-filling, where a spin-density wave instability is
leading, the compressibility is suppressed at low energy scales,
as expected for a system with a charge gap at or near the
chemical potential.
The homogeneous spin susceptibility remains finite near the
spin-density wave instability.
By contrast, further away from half-filling in the regime where
the $d$-wave pairing instability is leading, the compressibility
diverges while the homogeneous spin susceptibility is suppressed.
A suppressed spin susceptibility is expected as a precursor of 
the spin gap opening in any spin singlet superconductor. 
Very close to the instability the spin susceptibility flows 
through zero to negative values, which implies that our 1-loop
calculation breaks down in this strong coupling regime.
A diverging compressibility would indicate a tendency towards
phase separation, but the increase of $\kappa$ sets in quite
abruptly only very close to the instability, where the renormalized 
couplings are already so large that the 1-loop results are 
not reliable any more. In any case the
large charge fluctuations indicated by $\kappa\,\,\, \to\,\,\, \infty$
\widetext
\narrowtext
\begin{figure}[t]
\centering\leavevmode\epsfbox{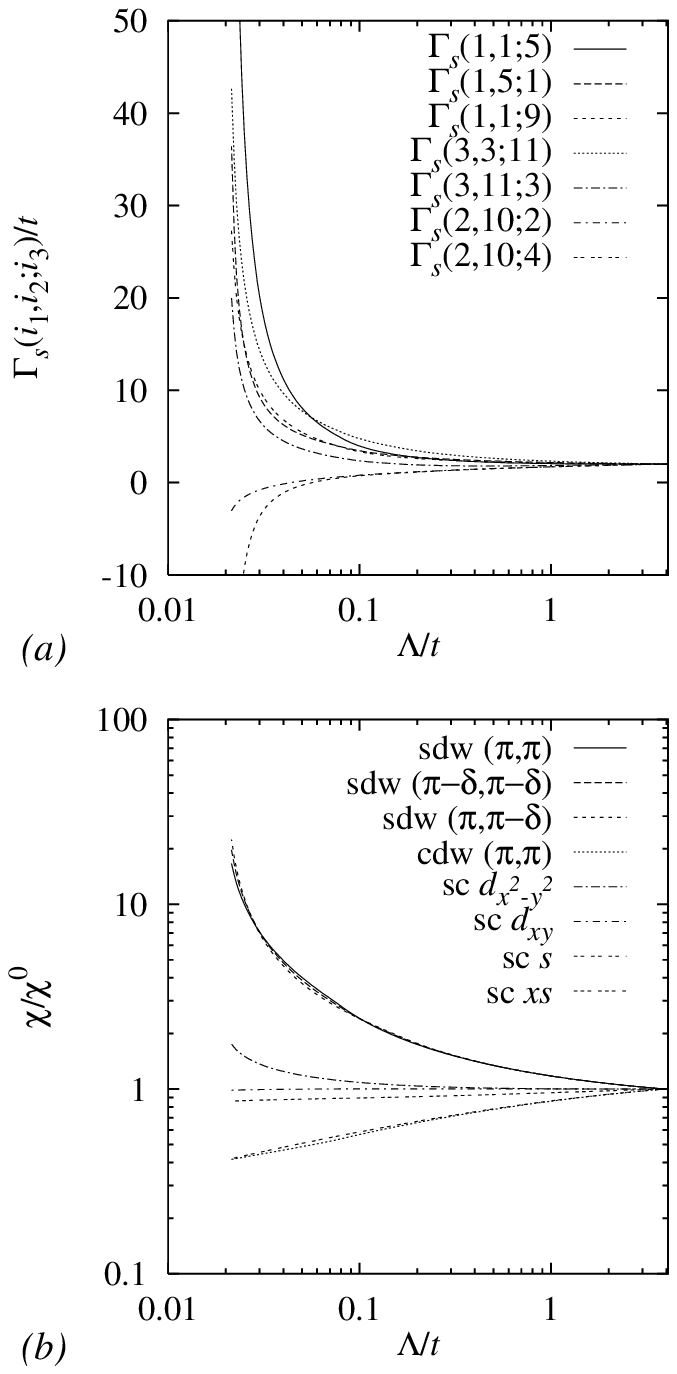}
\caption{Same as Fig.\ 6 for $U = t$, $t' = -0.01$ and
 $\mu = 4t'$.}
\label{fig11}
\end{figure}
\begin{figure}[t]
\centering\leavevmode\epsfbox{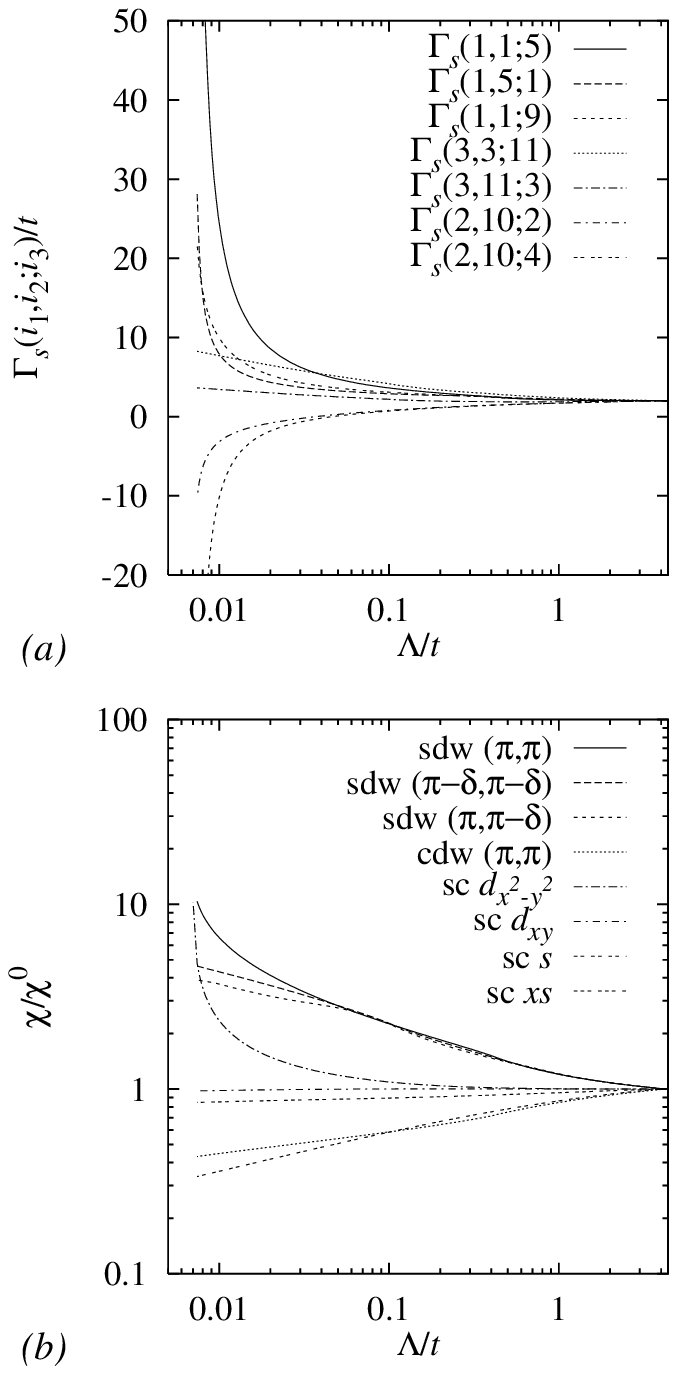}
\caption{Same as Fig.\ 6 for $U = t$, $t' = -0.05$ and
 $\mu = 4t'$.}
\label{fig12}
\end{figure}
\widetext
\narrowtext
would 
only be a consequence of the pairing instability of the Hubbard 
model, not a driving mechanism, since the pairing correlations
appear already at a higher energy scale. 

Results for the flow of the vertex function and susceptibilities for
$t'<0$ and $\mu = \eps_{vH} = 4t'$ are shown 
in Figs.\ 11 and 12, with $t' = -0.01$ and $t' = -0.05$, 
respectively. The corresponding Fermi surfaces touch the
saddle points at $(\pi,0)$ and $(0,\pi)$.
In the first case the density is $n=0.992$ and in the second one
$n=0.959$.
For the bare interaction we have chosen $U = t$ as before.
The major difference with respect to the perfect nesting case
$t'=0$ is that now the umklapp processes near the diagonal of
the Brillouin zone are much less enhanced at low energy
scales, such that scattering processes with momenta near the 
van Hove points $(\pi,0)$ and $(0,\pi)$ become most prominent.
In this situation the simple scaling approaches which 
concentrated exclusively on the van Hove points \cite{LMP,FRS} 
provide already a useful qualitative picture of the important
effective interactions and their renormalization.
 Antiferromagnetic correlations are now mostly driven by 
umklapp processes from $(\pi,0)$ to $(0,\pi)$ and vice versa 
which, due to the equivalence of the points $(\pi,0) = (-\pi,0)$
and $(0,\pi) = (0,-\pi)$ in the Brillouin zone, can also be 
viewed as Cooper processes. 
Indeed these processes are also responsible for $d$-wave pairing 
correlations.

For the parameters chosen in Fig.\ 11 antiferromagnetic 
correlations dominate over pairing. 
The incommensurate susceptibility with $\bq=(\pi,\pi-\delta)$ 
is a bit larger than the other incommensurate candidate and 
the commensurate antiferromagnetic susceptibility.
Note that there may be other still larger incommensurate 
susceptibilities among those not computed here. We have merely
investigated two (frequently discussed) incommensurate 
spin susceptibilities out of a variety of infinitely many 
candidates. Moving further away from half-filling, as in Fig.\ 12, one 
finds again dominant pairing susceptibilities, with $d_{x^2-y^2}$
symmetry in each case.

The phase diagram in the $(t',U)$-plane with $\mu = 4t' \leq$
\begin{figure}
\centering\leavevmode\epsfbox{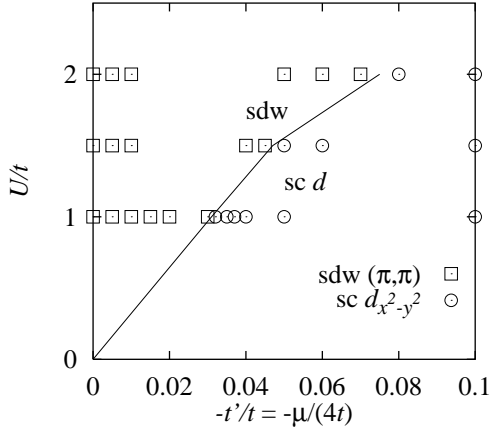}
\caption{The $(t',U)$ phase diagram for $\mu = 4t' \leq 0$.}
\label{fig13}
\end{figure}
\begin{figure}
\centering\leavevmode\epsfbox{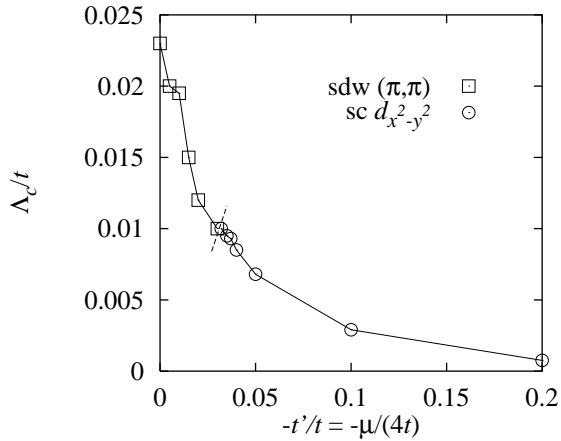}
\caption{The critical energy scale $\Lam_c$ as a function of
 $t'$ for $\mu = 4t' < 0$ and $U = t$; the short dotted line
 separates the spin density regime from the $d$-wave pairing
 regime.}
\label{fig14}
\end{figure}
\noindent
$ 0$ 
is shown in Fig.\ 13. 
Note that the chemical potential is
always situated 
at the van Hove singularity here and the density decreases away from
half-filling with increasing $|t'|$.
Since we have no good guess for the optimal density dependence of the 
incommensurability vector for $t' \neq 0$ we have not distinguished
different spin density waves in Fig.\ 13.
The behavior of $\Lam_c$ as a function of $t'<0$ with $\mu = 4t'$
and $U = t$ is shown in Fig.\ 14. 

The decrease of $\Lam_c$ with
increasing $|\mu|$ (and thus increasing doping) is slower here
than in Fig.\ 9, since the Fermi level remains on the van Hove
singularity such that only the importance of nesting is weakened
under doping.

All the numerical results discussed above have been obtained by 
projecting momentum variables of the vertex function on 16 points 
on the Fermi surface as shown in Fig.\ 5. 
To see how much these results may be modified in a more refined
projection scheme, we have computed the flow for some 
typical model parameters with a projection 
\begin{figure}
\centering\leavevmode\epsfbox{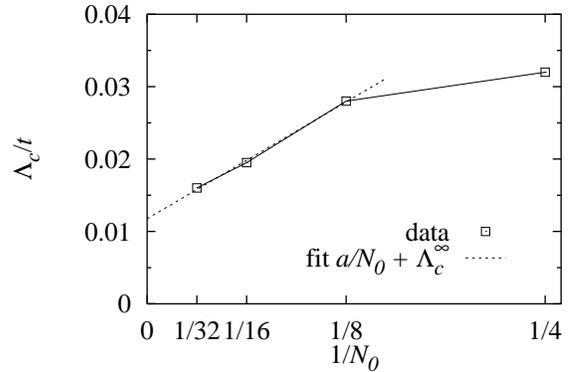}
\caption{The critical energy scale $\Lam_c$ as a function of
 the number of discretization points $N_0$ on the Fermi surface,
 for $U = t$, $t' = -0.01t$ and $\mu = 4t'$.}
\label{fig15}
\end{figure}
\noindent
on 32 points on the
Fermi surface, and also with a projection on 16 Fermi surface
points and 16 additional points on the van Hove surface.
It turned out that these refinements, which increase the 
computational effort considerablely, lead only to a moderate 
reduction of the critical energy scale, without changing the 
qualitative behavior of the vertex function and susceptibilities.
In Fig.\ 15 we show the dependence of the critical scale $\Lam_c$
as a function of the inverse number of discretization points
$N_0$ on the Fermi surface for $N_0 = 4,8,16,32$ and a fixed
choice of model parameters. 
We see that the critical energy scale obtained
from a discretization with 16 points has already the right
order of magnitude.

\section{Conclusion}
In summary, we have shown that the renormalization group method
developed by Salmhofer \cite{Sal1} with our extension for the
computation of susceptibilities can be used as a systematic tool 
for detecting instabilities in a weakly interacting Fermi system
with several coupled infrared singularities.
Such an RG analysis is completely unbiased.
The selection of retained Feynman diagrams is dictated by the 
weak coupling expansion and can be systematically improved
by including higher orders in a loop expansion.

Evaluating the flow equations on 1-loop level for the 2D Hubbard
model we have found antiferromagnetic instabilities close to
half-filling and dominant superconducting instabilities with 
$d_{x^2-y^2}$ symmetry at smaller densities (still near 
half-filling).
Incommensurate spin structures can be favorable in the 
antiferromagnetic regime near half-filling.

The critical energy scale $\Lam_c$ where vertex functions and
susceptibilities diverge vanishes exponentially as $U \to 0$,
but becomes sizable already for relatively weak coupling strengths
(compared to the band width), even in the superconducting regime. 
The appearance of strong pairing correlations with $d_{x^2-y^2}$ 
symmetry in the 2D Hubbard model at physically interesting energy 
scales is thus well established at weak coupling.
The flow of the vertex function and susceptibilities clearly shows 
that the pairing instability is driven by short-range 
antiferromagnetic correlations in the system.
This supports earlier ideas and numerical results (for finite
systems) suggesting $d$-wave superconductivity driven by 
antiferromagnetic correlations in the Hubbard model \cite{Sca}.
Note that $\Lam_c$ must not be interpreted as a transition 
temperature for antiferromagnetism or superconductivity, but
rather as an energy scale where bound particle-particle or 
particle-hole pairs are formed. A Kosterlitz-Thouless transition
to a superconducting state may occur at a lower energy scale
while antiferromagnetic order is of course possible only in the 
ground state of a two dimensional system with spin-rotation 
invariance. 

We finally outline some interesting extensions of the present
work for the future: \\
i) {\em Non-local interactions:}\/ Non-local interactions may
play an important role even though they are usually much smaller
than the local (Hubbard) interaction. They affect the RG flow
via a different initial condition for the vertex function and
can thus be taken into account very easily. \\
ii) {\em Fermi surface instabilities:}\/ The Fermi surface is
generally deformed by interactions. Computing a susceptibility
for Fermi surface deformations from the RG flow one finds
that deformations breaking the discrete square lattice symmetry
may occur \cite{HM2}. \\
iii) {\em Self-energy effects:}\/ It will be interesting to
compute self-energy contributions and see how they affect the 
instabilities. The numerical effort for this is small on 1-loop
level, but also a 2-loop calculation seems feasible.
Kishine and Yonemitsu \cite{KY} have recently computed the 
renormalization of the quasi-particle weight on 2-loop level for 
two flat Fermi surface pieces, but the feedback of self-energy
effects on instabilities has not yet been treated.

\vspace{4mm}
\noindent{\bf Acknowledgments:} \\
We are very grateful to Manfred Salmhofer for explaining to us
his renormalization group scheme before publication and giving 
us several useful hints during the initial stage of this work. 
We would also like to thank Maurice Rice, Eugene Trubo\-witz 
and Victor Yakovenko for valuable discussions. 
This work has been supported by the Deutsche 
Forschungsgemeinschaft under Contract No.\ Me 1255/4-1,2.
\appendix
\section{Non-interacting susceptibilities}
Here we show results for the flow of the non-interacting 
susceptibilities $\chi_0^{\Lam}$ for the choices of $t'$ and
$\mu$  corresponding to those in Fig.\ 6, 7, 11 and 12.
The reader may thus estimate the absolute scale of $\chi^\Lambda$ by
multiplying $\chi_0^{\Lam}$ with the results for the ratios
$\chi^{\Lam}/\chi_0^{\Lam}$ in Sec.\ III.
\begin{figure}
\centering\leavevmode\epsfbox{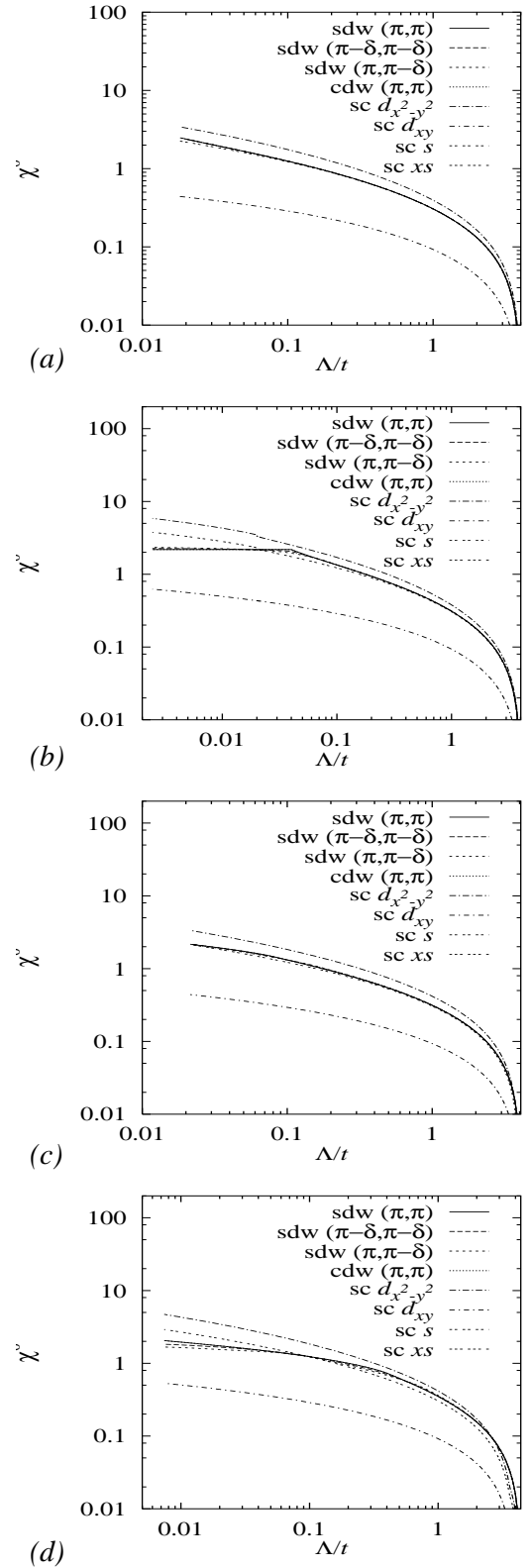}
\vspace{1mm}
\caption{Free susceptibilities for (a) $t'=0$ and $\mu=-0.005t$, (b)
$t'=0$ and $\mu=-0.02t$, (c) $t'=-0.01t$ and $\mu=4t'$ and (d) for
$t'=-0.05t$ and $\mu=4t'$, corresponding to the examples in Fig.\ 6,
7, 11 and 12, respectively.}
\label{figA}
\end{figure}
The incommensurate spin-density susceptibilities and the
charge-density susceptibility lie too close together to be always
individually seen. The extended $s$-wave pairing susceptibility is of
the order of $10^{-4}$ and therefore out of scale.


\widetext

\end{document}